\documentstyle[12pt,epsfig]{article}
\def \medio  {\baselineskip= 1.5 \normalbaselineskip}

\newcommand{\titul}[1] {\begin{center}{\large\bf #1 } \end{center}\vskip 1.cm}
\newcommand{\autor}[1] {\begin {center} {\large \lineskip .5em #1 }
                        \end   {center} }
\newcommand{\lugar}[1] {\begin{center} {\it #1} \end{center}}
\newcommand{\abstr}[1] {{\begin{center} \vskip .5cm {\bf Abstract
                        \vspace{0pt}} \end{center}}\begin{quote} #1
                        \end{quote}}

\begin{document}
\begin{titlepage}


\begin{flushright} {} \end{flushright}

\vspace{3.cm}

\titul{ 
Charged Current Neutrino Cross Section and
\\ Tau Energy Loss at Ultra-High Energies}
\autor{
N. Armesto\footnote{nestor@fpaxp1.usc.es}, 
C. Merino\footnote{merino@fpaxp1.usc.es}, 
G. Parente\footnote{gonzalo@fpaxp1.usc.es}, and 
E. Zas\footnote{zas@fpaxp1.usc.es}
}
\lugar{Departamento de F\'\i sica de Part\'\i culas $\&$\\ 
Instituto Galego de F\'\i sica de Altas Enerx\'\i as\\
Universidade de Santiago de Compostela\\
15706 Santiago de Compostela, Spain}
\abstr{
\medio

We evaluate both the tau lepton energy loss produced by photonuclear
interactions and the neutrino charged current cross section
at ultra-high energies,
relevant to neutrino bounds with Earth-skimming tau neutrinos, 
using different theoretical and phenomenological models for nucleon and
nucleus structure functions. 
The theoretical uncertainty is estimated by taking different extrapolations 
of the structure function $F_2$ to very low values of $x$,
in the low and moderate $Q^2$ range for
the tau lepton interaction and at high $Q^2$ for the neutrino-nucleus inelastic
cross section. It is at these extremely low values of $x$ where nuclear
shadowing and parton saturation
effects are unknown and could be stronger than usually considered.
For tau and neutrino energies $E=10^{9}$ GeV we find uncertainties of a factor
4 for the tau energy loss
and of a factor 2 for the charged current neutrino-nucleus cross section. 

}
\end{titlepage}
\newpage

\pagestyle{plain}
\medio
\section{Introduction} \indent


The detection of high energy neutrinos is one of the most important challenges 
in Astroparticle Physics. 
Conventional neutrino detectors exploit the long range of muons produced
by muon neutrino charged current (CC) interactions \cite{Halzen}.
With the discovery of neutrino flavor oscillations
it has been realized that also tau neutrinos reach the Earth
in spite of being  heavily suppressed in all postulated production mechanisms. 
The possibility to search for tau neutrinos 
by looking for tau leptons that exit the Earth, 
Earth-skimming neutrinos, has been 
shown to be particularly advantageous to detect neutrinos of energies 
in the EeV range~\cite{Fargion,Bertou}. 
The short lifetime of the tau lepton originated in the neutrino
charged current interaction allows the tau to decay in flight while
still close to the Earth surface producing an outcoming air shower in
principle detectable by both fluorescence telescopes and air shower arrays
\cite{EZas}. This same channel yields negligible contributions for
other neutrino flavors. 
The sensitivity to tau neutrinos through the Earth-skimming channel
directly depends both on the neutrino charged current cross section
and on the tau  range (the energy loss)
which determine the amount of 
matter with which the neutrino has to interact to produce an emerging
tau~\cite{ICRCAuger_neutrino1,ICRCAuger_neutrino2}.
While the energy loss for muons is 
shared by roughly equivalent contributions from
pair production, bremsstrahlung and photonuclear interactions, 
for tau leptons of energies above $E = 10^7$ GeV, 
photonuclear interactions
(i.e. lepton-nucleus inelastic interactions dominated by small values of $Q^2$) are responsible 
for the largest and the most uncertain contribution~\cite{Dutta2001,Bugaev2004,Aramo2005}. 

Both the neutrino cross section and the tau photonuclear energy loss 
are calculated from theory using
structure functions which carry the information of the nucleon and nucleus
structure.
In order to study the uncertainties in the calculation of Earth-skimming
neutrinos the same structure functions should be consistently used for both 
processes due to their strong correlation in the resulting tau flux.
Unfortunately this is not possible since the kinematical
$Q^2$ (minus the squared momentum transfer) and
Bjorken-$x$ ranges that contribute to these processes are quite different, 
specially at EeV energies, and the available parameterizations are not 
entirely adequate to describe both ranges simultaneously.  

The $Q^2$ scale that contributes to the tau energy loss, dominated by photon 
exchange, is low and moderate $Q^2$ at very low $x$,
where perturbative and non perturbative QCD
effects are mixed. 
The CC neutrino cross section is produced by $W$-boson 
exchange that sets the relevant scale of $Q^2$ to
values up to $M_W^2~$ at low $x$, a region
where perturbative QCD is expected to work. 
In both cases the relevant $x$ range
lies well outside the regions where structure functions are
measured, so one has to rely on extrapolations which contain significant
uncertainties.


The charged current neutrino cross section is usually calculated using parton 
distribution 
functions which are evolved according to perturbative QCD predictions. 
A number of alternative parameterizations exist, some of
which allow extrapolation of the uncertainties in the fitted parameters
as a mean to explore some of the uncertainties associated to the
calculation. In the case of photonuclear processes
existing predictions at high energy arise basically from two independent 
approaches, the Generalized Vector Dominance (GVD) model
and Regge-like models. 

In this article we study the tau energy loss
(see also Ref.~\cite{nosoICRC07})
and the neutrino-nucleus cross section. Both quantities have direct
implications for high energy 
neutrino detection, in particular for Earth-skimming tau 
neutrinos. 
Due to the large uncertainties in the existing models, the fact that
none of them covers simultaneously the kinematical region relevant for
both quantities, and the need of consistency in both calculations, we use
and extend available models with the aim of estimating the theoretical
uncertainty by considering extreme results.
In this way, in the frame of the most relevant models, we cover the
range of possible scenarios for the extrapolation of
structure functions to the relevant $x$ and $Q^2$ range.
Two important effects to be taken into account in this extrapolation of
the structure functions are nuclear shadowing corrections and saturation
due to partonic screening. 
Nuclear corrections~\cite{nuclear} are deviations from the naive picture in
which the nucleus is treated as an incoherent sum of nucleons. 
Saturation~\cite{saturacion} accounts for the fact
that the structure functions cannot rise indefinitely as $x$ goes to zero.
Saturation effects may be included in nuclear corrections but are also present 
in the nucleon structure functions 
(although for smaller values of $x$ and/or $Q^2$). 
In addition to existing calculations, a new computation of
the tau energy loss and the neutrino-nucleon cross
section based on saturation physics~\cite{ASW} is also presented in this work.

The result of the present analysis is an uncertainty band 
for both the tau-lepton energy loss by photonuclear interactions
and the CC neutrino-nucleus cross section. 
Understanding and minimizing the uncertainties 
in these two calculations must be considered an important priority for 
high energy neutrino astrophysics.

\section{The photonuclear tau energy loss}
\indent

The average energy loss per unit depth, $X$, of taus 
is conveniently represented by:
\begin{eqnarray}
- \left<\frac{dE}{dX}\right> = a(E) + b(E) E \; ,
\end{eqnarray}
where $a(E)$ is due to ionization and $b(E)$ is the sum of
fractional losses due to e$^+$e$^-$ pair production,
bremsstrahlung, and photonuclear interactions. 
The parameter $a(E)$ is nearly constant and the term $b(E) E$ 
dominates the energy loss above a critical energy that for tau 
leptons is of a few TeV. 
The electromagnetic contribution to the energy loss, mainly due to pair 
production and bremsstrahlung,  is well under control,
while  the photonuclear interaction which dominates for tau energies exceeding
$E=10^7$~GeV is affected by relatively large uncertainties.  

The contribution to $b(E)$ from photonuclear
interactions is obtained by integration of the 
lepton-nucleus differential cross section, $d\sigma^{lA}/dy$: 
\begin{eqnarray}
b(E) = 
\frac{N_A}{A} \int dy \; y \int dQ^2 \frac{d\sigma^{lA}}{dQ^2 dy} \; ,
\end{eqnarray}
where $N_A$ is Avogadro's number, $A$ the mass number,
and $y$ the fraction of energy 
lost by the lepton in the interaction. 

For the lepton-nucleus differential cross section we consider the general 
expression for virtual photon exchange in terms of structure functions:
\begin{eqnarray}
\frac{d\sigma^{lA}}{d Q^2 dy} = \frac{4 \pi \alpha^2}{Q^4}\frac{F_2^A}{y}
\left[ 1 - y - \frac{Q^2}{4 E^2} + \left(1-2\frac{m_l^2}{Q^2}\right)
\frac{y^2 + Q^2/E^2}{2(1+R^A)} \right] \; ,
\end{eqnarray}
where $E$ is the lepton energy in the lab frame, 
$m_l$ the lepton mass, and $\alpha$ the fine structure constant.
$F_2^A$ is the structure function $F_2$ for a nuclear
target $A$ which is found 
to be different from the mere superposition of $A$ free nucleon
structure functions $F_2^p$~\cite{nuclear}. 
$R^A$ is the ratio of the longitudinal to transverse structure functions
which gives a small contribution to the cross section
\cite{Dutta2001} and is neglected for clarity of the discussion below. 
The variables $x$, $y$ and $Q^2$ are related by kinematics through $Q^2=2MExy$, 
and both $F_2$ and $R$ are functions of $x$ and $Q^2$. 
The contribution to the tau energy loss from neutral current and
$\gamma$-$Z$ interference interactions was estimated to be small
\cite{BM2002} and is also neglected. 

The limits in the double integral of Eq. (2) are well established:
\begin{eqnarray}
Q^2_{min}=\frac{y^2 m_l^2 }{1-y} \; , \;\;\;\;\;\;\;\;\;\;
Q^2_{max}=2m_pEy -2m_{\pi}m_p-m_{\pi}^2 \; , \\
y_{min}=\frac{2 m_{\pi} m_p+ m_{\pi}^2}{2 m_p E} \;, \;\;\;\;\;\;\;\;\;
y_{max}=1-\frac{m_l}{E} \; ,
\end{eqnarray}
where $m_p$ and $m_{\pi}$ are the proton and pion mass, respectively.

The predictions of the photonuclear interaction cross section
in the GVD Model~\cite{BB} (BB) and in its extension to higher energies
by including a perturbative component based on the color dipole
model~\cite{BS2003} (BS),
have been widely used to explore muon and tau lepton
propagation in matter (see for instance~\cite{Lipari,Bugaev2004,Aramo2005}
and references therein). 

The calculations
in which the $F_2$ structure function
is given by a phenomenological parameterization of data based
on Regge Theory appear in  Refs.~\cite{Dutta2001} (DRSS),
\cite{BM2002} (BM), \cite{KLS2005} (KLS),
and \cite{Petrukhin} (PT).
%
%
For the proton structure function, $F_2^p$, DRSS (see also
Ref. \cite{Dutta2005}) uses the ALLM 
model \cite{ALLM}, while BM and KLS both consider the CKMT
model~\cite{CKMT} at low $Q^2$ matched at high $Q^2$ to 
perturbative QCD predictions 
based on different parameterizations of 
parton distribution functions, and PT uses the proton structure function
of Ref.~\cite{F2petrukhin}. 
The $F_2^p$ structure function is  shown in Figs.~\ref{figF2a} 
and \ref{figF2b}, together with the HERA data at the lowest
measured $x$ values at different $Q^2$.

\begin{figure*}[t]
\vskip5.cm 
\includegraphics[width=13.0cm]{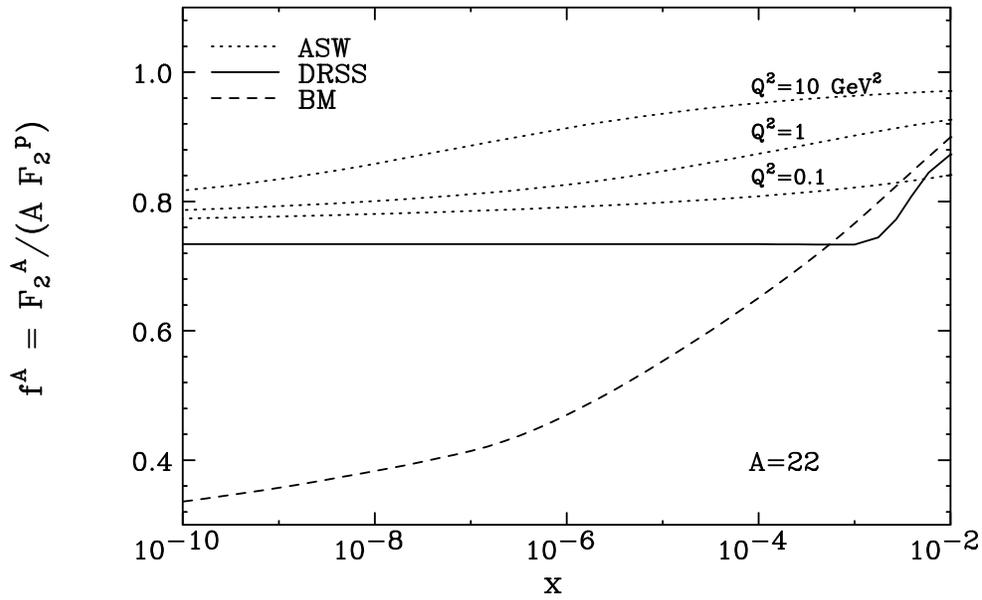}
\caption{The nuclear correction factor $f^A$ as a function of $x$. }
\vskip5.cm
\label{figNuc}
\end{figure*}
\begin{figure*}[t]
 \includegraphics[width=13.0cm]{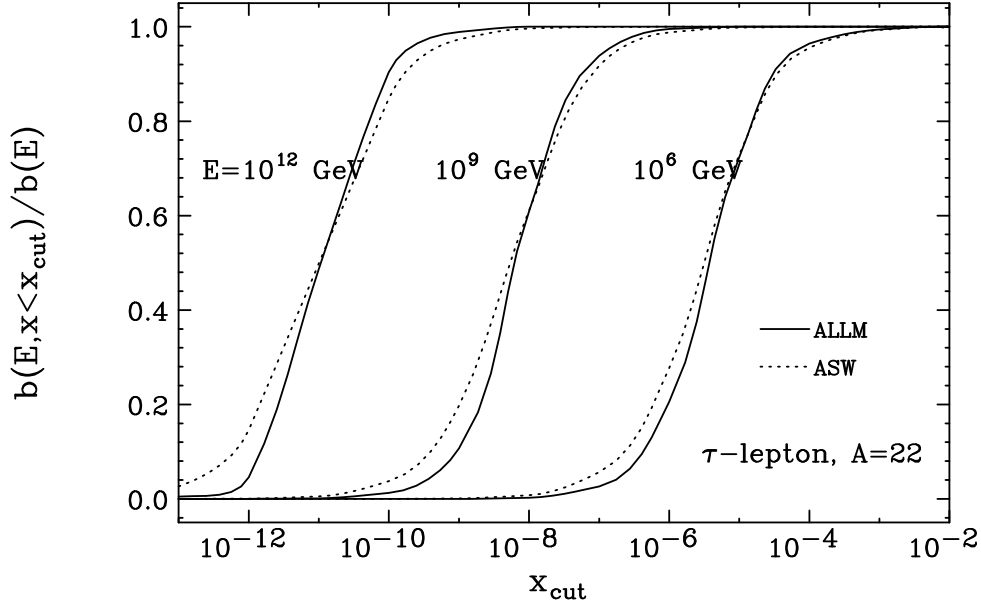}
\caption{The relative contribution of $x<x_{cut}$
to the photonuclear energy loss rate, $b(E)$.}
\label{figrelx}
 \end{figure*}

In DRSS, BM, and KLS calculations the nuclear structure function is
related to the proton structure function through $F_2^A= f^A A F_2^p$.
At high energy only the low $x$ behavior 
of the nuclear correction factor $f^A$ is relevant to the
calculation of $b(E)$, as we will show below (see Fig.~\ref{figrelx}). 
In the DRSS calculation the low $x$ behavior of $f^A$ freezes at the value
$f^A=A^{-0.1}$ for  $x<0.0014$ ($\sim 0.73$ for standard rock, $A=22$),
while in the BM (and KLS) calculations 
$f^A$ reaches a  maximal asymptotic regime
$f^A=A^{-1/3}$ ($\sim 0.36$ for $A=22$) at much lower $x$
(see Fig.~\ref{figNuc}). 
Both DRSS and BM nuclear corrections are $Q^2$-independent.

\newcommand{\qs}{Q_{\rm sat}}
\newcommand{\qsa}{Q_{\rm sat, A}}

In addition to the existing calculations
we present a new computation of the photonuclear tau energy loss  
using the  results of Ref. \cite{ASW} (ASW)
which are based on the geometric scaling property \cite{Stasto} that 
all data on $\sigma^{\gamma^* p}$ and on
$\sigma^{\gamma^* A}$ lie on a single universal curve
in terms of the scaling variable $\tau=Q^2/Q^2_{sat}$
whose form is inspired in saturation physics
(the detailed expressions leading to the ASW $F_2$ structure function
are given in the Appendix). 
The ASW $F_2$ structure function for the proton case
is plotted in Figs.~\ref{figF2a} and~\ref{figF2b}
(for $x<0.01$ where this parameterization is expected to be valid). 
The ASW structure function $F_2$ contains mild nuclear corrections at low $x$
when compared with DRSS and BM nuclear corrections (see Fig.~\ref{figNuc}).
Nuclear corrections in ASW depend on $Q^2$.

\begin{figure*}[t]
\includegraphics[width=13.0cm]{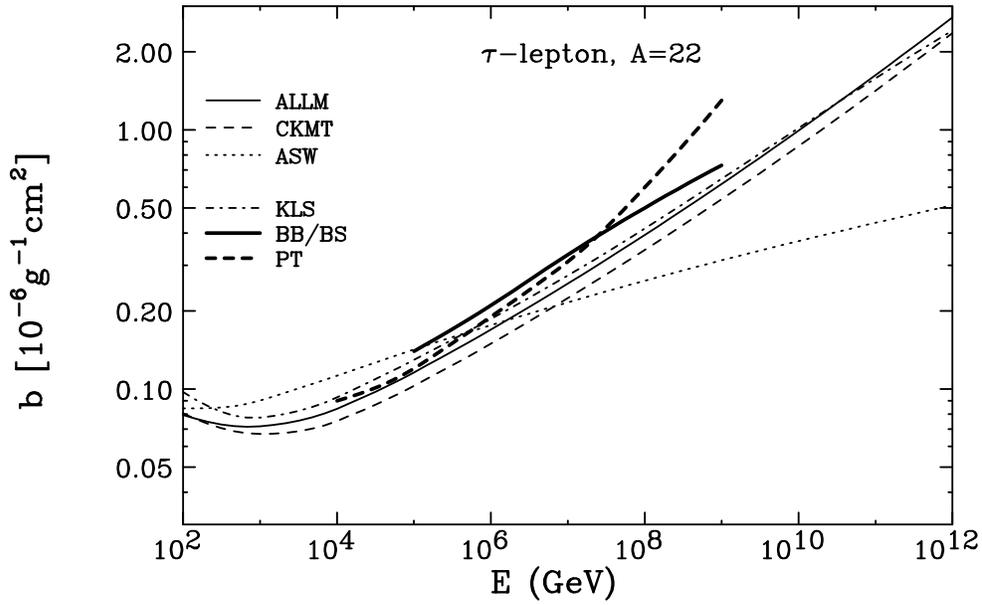}
 \caption{The photonuclear energy loss rate, $b(E)$, computed in different
models.}
\label{figloss}
 \end{figure*}

The photonuclear contributions to $b(E)$ computed
(for standard rock $A=22$ throughout all this paper)
with ALLM and with CKMT structure functions, and the same nuclear
corrections \cite{Dutta2001}, give very close results
(see Fig.~\ref{figloss}). 
Although ALLM and CKMT parameterizations share
a common theoretical base, with a reggeon and a pomeron component,
and they are fitted to the same data sets, 
ALLM systematically lies above CKMT at low $x$  (see Fig.~\ref{figF2b}),
which accounts for the difference in $b(E)$ observed in Fig.~\ref{figloss}.

The lowest values of $b(E)$ at high energies is
obtained with the ASW structure functions.
Though the ASW structure function $F_2$ contains mild nuclear
corrections at low $x$, saturation effects at the nucleon
level are rather strong and limit the rise of $b(E)$ with energy
as observed in Fig.~\ref{figloss}.
For energies below $E=10^6$~GeV the result
from the ASW structure function is higher than those
from ALLM or CKMT (see Fig.~\ref{figloss}). This is because at low $Q^2$
the ASW structure function
is significantly higher for the region
$10^{-6}<x<10^{-3}$ (see Fig.~\ref{figF2b}) which is the relevant range
for energies below $E=10^6$ GeV, as it can be deduced from Fig.~\ref{figrelx}. 
Thus the saturation-based ASW prediction lowers
the energy loss rate $b(E)$ with respect to the already
existing predictions by a factor 2 at  $E=10^9$~GeV, and
by a factor even larger at higher energies.

The BB/BS calculation gives
the largest of the predicted energy loss rates up to energies of the order 
$E=10^7$~GeV. Above this scale the PT result exceeds all other
existing predictions by at least a factor 2 already at $E=10^9$~GeV 
(i.e. a factor 4 with respect the ASW prediction, see Fig.~\ref{figloss}). 
Thus the PT prediction can be considered as an estimate of the upper limit 
of the tau energy loss at UHE. 

\begin{figure*}[t]
\includegraphics[width=13.0cm]{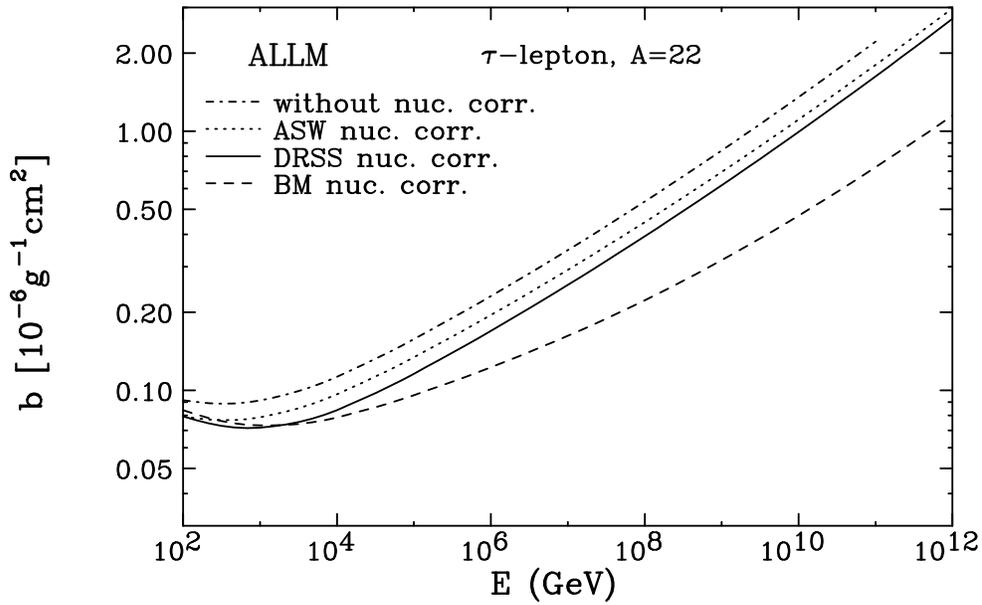}
\caption{The effect of nuclear corrections on the photonuclear energy
loss rate, $b(E)$.}
\label{figlossnuc}
 \end{figure*}
\begin{figure*}[b]
 \includegraphics[width=13.0cm]{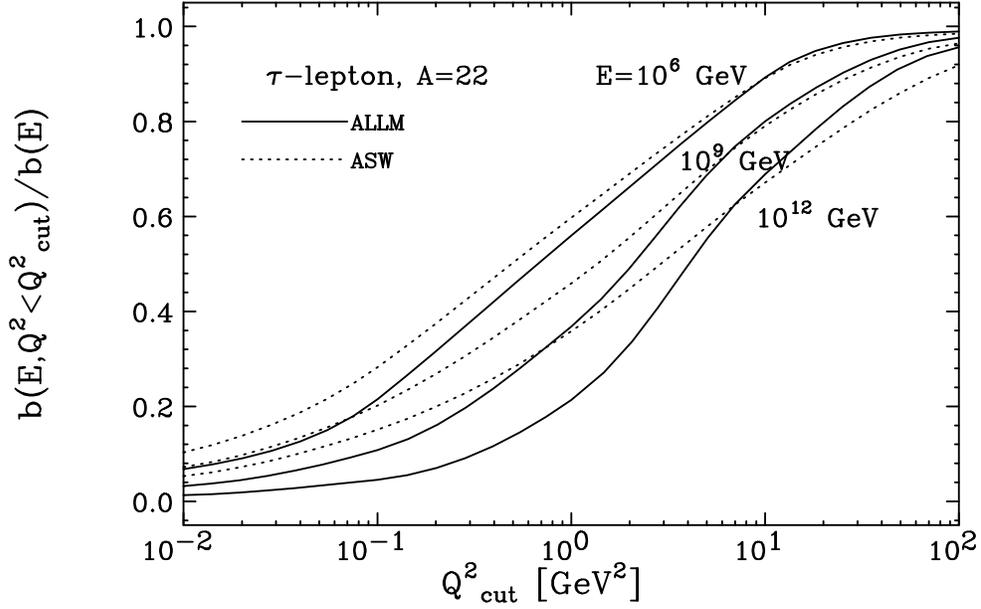}
\caption{The relative contribution  of $Q^2<Q^2_{cut}$
to the photonuclear energy loss rate, $b(E)$.}
\label{figrelq2} 
 \end{figure*}

Much of the uncertainty in the
tau energy loss is actually due to nuclear effects. 
The choice of nuclear corrections from Ref. \cite{Dutta2001}, 
Ref. \cite{BM2002}, or from Ref. \cite{ASW} (see Fig.~\ref{figNuc}), 
translates  into differences in the calculated value of $b(E)$ 
(using the ALLM  structure function) by a factor rising from 1.5 to 2.5 
as the tau energy increases in the range $E=10^{6}$-$10^{9}$~GeV
(see Fig.~\ref{figlossnuc}). This energy range corresponds to the region 
of very low $x$ where differences in the nuclear correction factor are large. 
In order to quantify how much different regions of $x$ and $Q^2$ 
contribute to $b(E)$, the dependence of $b(E)$ on the maximum value 
of $x$ and on the maximum value of $Q^2$ considered in the integration 
is shown in Figs.~ \ref{figrelx} and~\ref{figrelq2}. 

\begin{figure*}[t]
 \includegraphics[width=13.0cm]{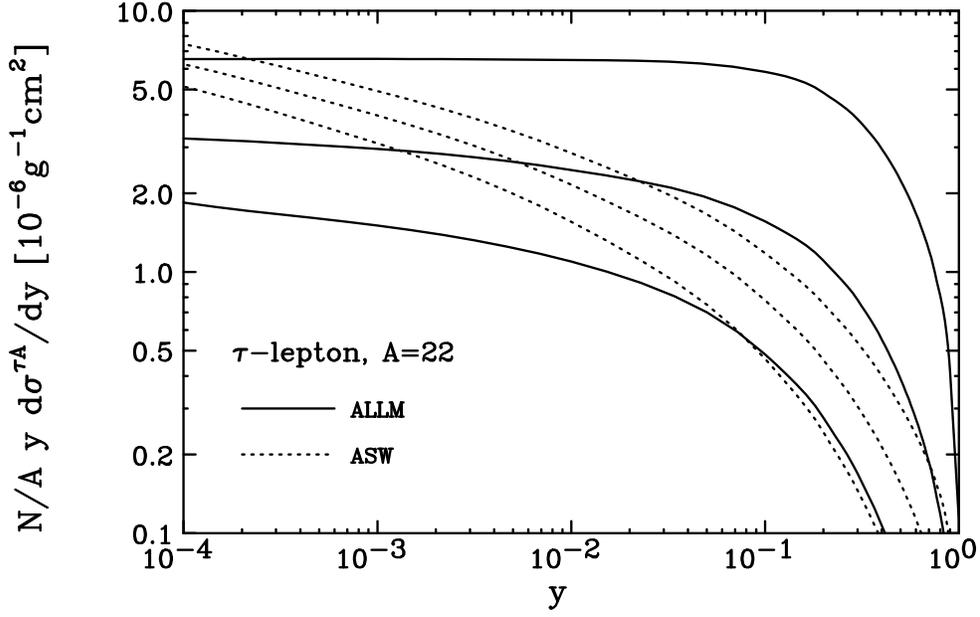}
 \caption{Spectrum of the tau energy loss by photonuclear interactions
for tau energies $E=10^{6}$, $10^{9}$, and $10^{12}$ GeV.}
\label{figdsig} 
 \end{figure*}
\begin{figure*}[t]
 \includegraphics[width=13.0cm]{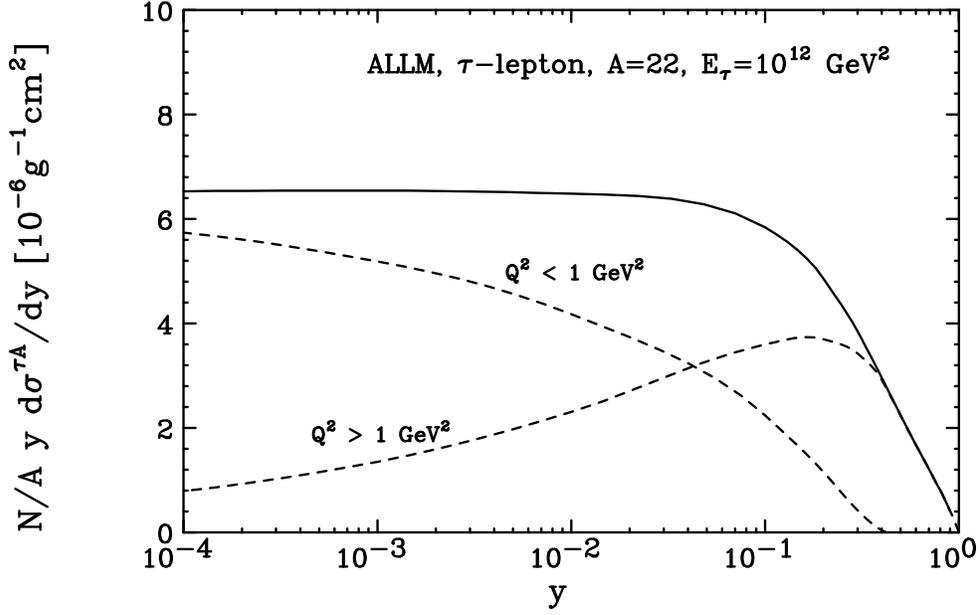}
 \caption{Spectrum of the tau energy loss by photonuclear interactions for
a tau of energy $E=10^{12}$ GeV. The contributions from values above
and below $Q^2=1$ GeV$^2$ are shown separately.}
\label{figdsig12} 
 \end{figure*}

The differential cross section $d\sigma^{\tau A}/dy$ is also
a relevant quantity 
for high energy neutrino detection as it enters the event rate convolutions 
together with the neutrino flux and the experimental acceptances. 
Indeed it has been shown that stochastic effects of the tau energy loss 
distribution have significant relevance in the prediction of emerging tau 
rates~\cite{Dutta2005}. 
The energy loss spectrum $y d\sigma^{\tau A}/dy$ obtained using 
both ALLM and ASW structure functions are compared in Fig.~\ref{figdsig}. 
Clearly, the energy loss spectrum calculated with 
ALLM is significantly harder than the one calculated with ASW. 
The contributions of moderate ($Q^2>1$) and low $Q^2$ ($Q^2<1$) 
(in a rough way corresponding respectively to hard and soft interactions)
are shown separately in Fig.~\ref{figdsig12} 
for the ALLM structure function. 


\section{The charged current neutrino cross section}
\indent

The absolute value of the cross section naturally has a direct impact on the 
sensitivity of experiments because the event rate is directly proportional to
it, but it also enters with opposite effect in the attenuation of the
neutrino beam as a function
of matter depth traversed, having much impact on the 
angular and energy distribution of the events. These two effects combine 
in the case of Earth-skimming tau neutrino interactions to play an important
role for the rate calculation. 
In addition to the tau lepton photonuclear cross section
we also study how the uncertainties in the
$F_2$ structure function at low $x$ 
affect the CC neutrino deep inelastic cross section.
Since in the more realistic expectations
\cite{ReportNestor} the nuclear corrections to the CC neutrino-nucleon
cross section decrease at low $x$
with increasing $Q^ 2$, becoming small at high $Q^ 2$~\cite{CastroPena},
we will neglect them in our calculations.

The CC DIS neutrino-nucleon cross section is expressed in terms of the
structure function $F_2$ as follows:
\begin{eqnarray}
\frac{d\sigma_{CC}^{\nu N}}{d Q^2 dy} = 
\frac{G_F^2}{4 \pi} \left(\frac{M_W^2}{M_W^2+Q^2}\right)^2
\frac{F_2^{\nu N}}{y} [ 1+(1-y)^2] \; ,
\end{eqnarray}
where $E$ is the neutrino energy and $y$ the fraction of energy
lost by the neutrino in the interaction. In this expression 
$F_L$ and $xF_3$ contributions are neglected since $F_L$ tends to zero 
as $Q^2$ rises and $xF_3$ deals basically with the valence partons 
which hardly contribute at the low $x$ values relevant for 
the cross section. 

In order to consistenly use the structure functions from
charged lepton interactions (as ALLM, CKMT, and ASW models) in 
neutrino interactions we must relate the electromagnetic and weak
structure functions.
The $F_2$ structure function for neutrino interaction is related 
to the $F_2$ structure function 
for charged lepton interactions by the ratio of the weak and electromagnetic 
couplings through $F_2^{\nu N}=18/5 \; F_2^{lN}$ (assuming a symmetric sea),
although the kinematical regions of the two processes are
different (low and moderate $Q^2\sim 0.01$-$10$ GeV$^2$ in the 
photonuclear case
and high $Q^2 \sim M_W^2$ in the high energy CC interaction).
Concerning the $x$ range the main contribution comes
from low $x$ in both cases, though $x$ values are
lower in the photonuclear case than in the CC interaction.
For the calculation of the neutrino-nucleon cross section at
high energies, we then use the structure function $F_2$ for charged lepton
interaction valid up to very low $x$ and high $Q^2$, instead
of following the standard approach based on parton densities.

\begin{figure*}[t]
 \includegraphics[width=13.0cm]{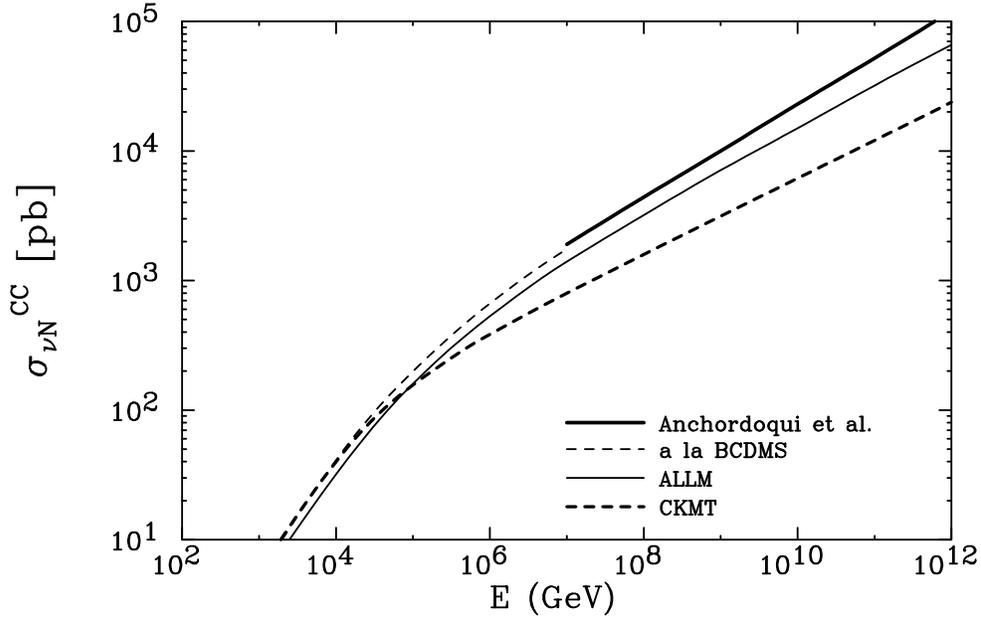}
 \caption{The neutrino-nucleon CC cross section as a function of the
neutrino energy,~$E$, from: 
ALLM, CKMT,  SMC {\it \`a} $la$ BCDMS structure functions,
and the parton density calculation
by Anchordoqui {\it et al.} }
\label{fignuxsection1} 
 \end{figure*}

The neutrino-nucleon cross sections from ALLM and CKMT structure functions
are presented in Fig.~\ref{fignuxsection1}. 
They are clearly
below predictions from  modern parton densities \cite{Anchordoqui}, 
since the ALLM parameterization is not consistent
with high $Q^2$ experimental points (see Fig.~\ref{figF2a}) and 
CKMT is not evolved to high $Q^2$, so we have not used them to discuss
the theoretical uncertainties in the estimation of
the CC neutrino-nucleon cross section.
Instead we have taken the parameterization of $F_2$
{\it \`a} $la$ BCDMS obtained by the SMC Collaboration~\cite{SMC}, which
correctly represents the existing
experimental data at high $Q^2$ (see Fig.~\ref{figF2a}) and provides
a smooth connection at neutrino energies around $E=10^7$ GeV with
the parton density prediction
of the CC neutrino-nucleon cross section \cite{Anchordoqui}
(see Fig.~\ref{fignuxsection1}).

We have performed three different extrapolations at low $x$ of the
$F_2$ parameterization given in Ref. \cite{SMC},
one following the ASW structure function, a second one from the
phenomenological parameterization fitting
low $x$ HERA data \cite{HERA}, and the third one which corresponds to
the double logarithmic approximation (DLA) in QCD from
Ref. \cite{KOPA} (KOPA) (ASW and KOPA structure functions are valid
at low $x$, $x<0.01$, i.e. at high energy).

\begin{figure*}[t]
 \includegraphics[width=13.0cm]{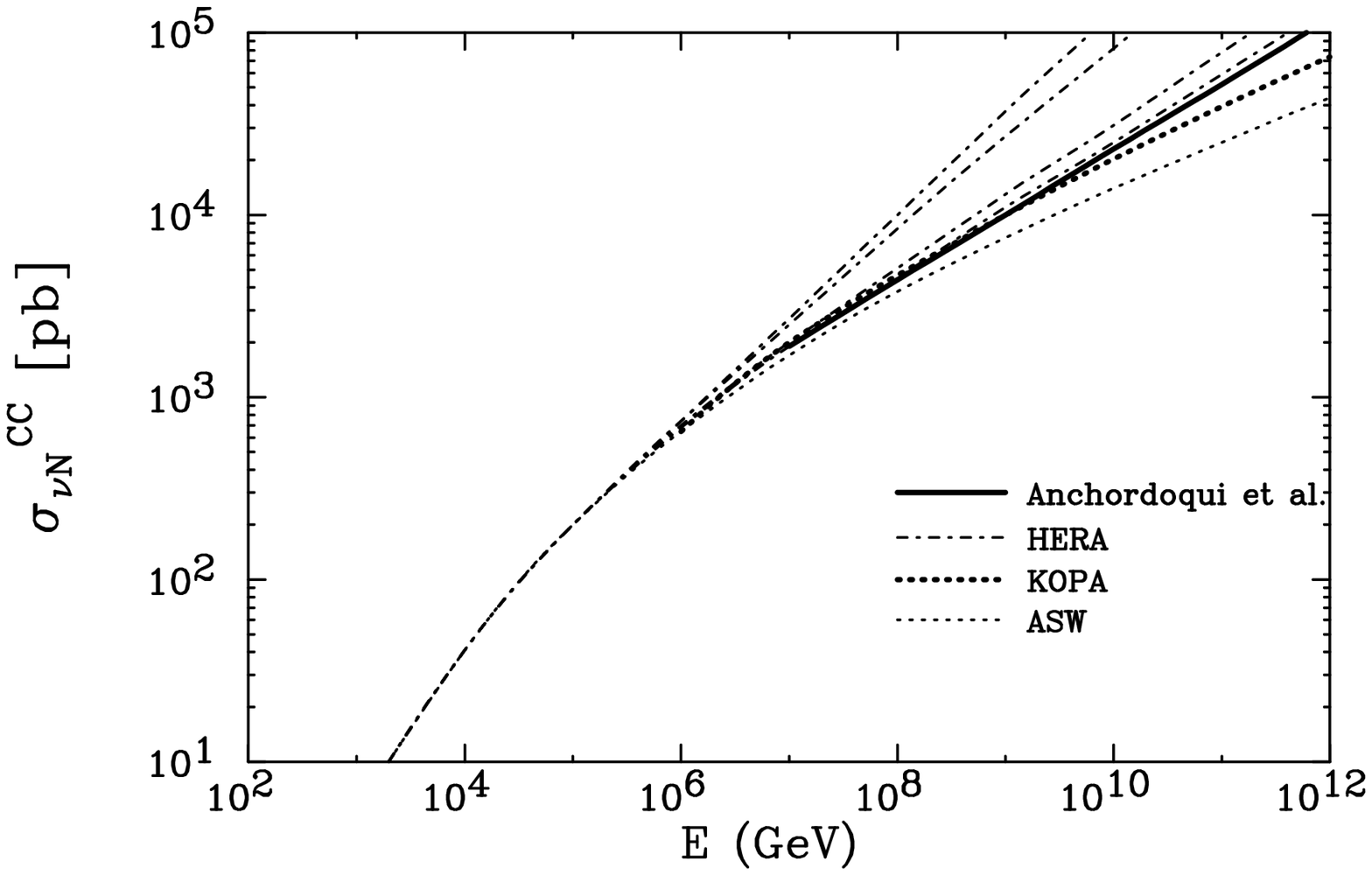}
 \caption{The neutrino-nucleon CC cross section as a function of the
neutrino energy,~$E$, from:
ASW (extrapolation based on saturation physics), KOPA (extrapolation based
on DLA QCD), HERA (phenomenological parameterization of HERA data), and
Anchordoqui {\it et al.} (parton density calculation).
}
\label{fignuxsection2} 
  \end{figure*}

In Fig.~\ref{fignuxsection2} the effect 
of taking the three different parameterizations of the structure function $F_2$
at low $x$ on the neutrino-nucleon cross section  is shown.
We see that in comparison with the prediction obtained with evolved QCD parton
densities of Ref.~\cite{Anchordoqui}, both KOPA 
(which corresponds to the DLA of perturbative QCD)
and ASW (which includes strong saturation
effects) estimations are below at high energies.

On the other hand the extrapolation of the HERA based parameterization
with the exponent 
$\lambda=0.0481 \ln(Q^2/0.292^2)$ ($F_2 \sim x^{-\lambda}$),
produces an extremely fast increase of the cross
section with energy (see the upper curve in  Fig.~\ref{fignuxsection2}),
since this exponent rises to values above $\lambda \sim 0.5$
when $Q^2$ becomes large. This raw extrapolation is in contradition with
perturbative calculations and we do not consider it for uncertainty estimates
as it is not physically motivated. Nevertheless it is considered here
to explicitely show its discrepancy with pQCD.
For the more realistic scenarios, when the rise of the exponent freezes to
smaller values $\lambda < 0.4$, our prediction supports the result 
obtained in the detailed analysis of Ref.~\cite{Anchordoqui}.
The curves are shown in  Fig.~\ref{fignuxsection2} (from up to down 
corresponding to  $\lambda$ frozen to $\lambda=0.50$, $0.40$, and $0.38$
respectively).
When considering only physically motivated extrapolations,
the theoretical uncertainty at $E=10^{9}$ GeV is a factor 2.

\vspace{-0.25cm}

\section{Conclusions}
\indent

We estimate the uncertainties coming from the extrapolations
of the existing models for proton and nucleus structure functions
for tau energy loss and for CC
neutrino-nucleon cross section.
Both calculations must be done consistently within the same model as
their effect on the tau flux produced by Earth-skimming neutrinos
is correlated.
The theoretical uncertainty in the tau energy loss is greater than 
that of the neutrino-nucleon CC cross section
because the $Q^2$ region contributing to the tau energy loss 
cross section is lower and so are the relevant values of Bjorken-$x$. 
In addition the structure functions conventionally used for the calculation 
of the tau energy loss are not suitable to be used in the high-$Q^2$ range 
which is relevant for the CC neutrino-nucleon interaction.
As a result systematic effects arising in the calculation of
a tau neutrino bound from Earth-skimming events due
to uncertainties in the structure functions turn out to be difficult
to evaluate.
Several extreme models allowed by extrapolation of structure 
functions have been explored in order to estimate ranges for these quantities. 

Below energies in the $E=10^7$~GeV range the highest prediction for the 
photonuclear contribution to tau energy loss, $b(E)$, is provided by the 
BB/BS calculation. Above this energy range the PT result exceeds all other 
considered predictions while the lowest calculation is obtained using the 
ASW structure functions. 
The difference between the two extreme predictions 
reaches a factor 4 at $E=10^9$~GeV and increases as the energy rises. 
The BB/BS, ALLM, and CKMT calculations agree within a 30~$\%$ and go
approximately parallel for all energies, which is an indication of a 
systematic normalization difference of the structure functions in each 
model.
The application of much stronger nuclear shadowing 
(than usually considered)  at low $x$ 
can lower the prediction of $b(E)$ with respect to the already 
existing calculations by a factor up to 2 at $E=10^9$~GeV.

In the case of the CC neutrino-nucleon cross section the importance of nuclear
effects at high energies is expected to be small~\cite{CastroPena}.
We have also considered saturation effects in the CC neutrino-nucleon 
cross section by using the structure function ASW. 
At $E=10^{10}$ GeV, the CC neutrino-nucleon cross section calculated with the
ASW structure function is found to be half of the pQCD calculation
with parton densities, in rough agreement with the evaluation of
saturation effects reported in Ref.~\cite{Kutak_Kwiecinski},
and also in Ref.~\cite{Henley:2005ms} where different parameterizations of
the dipole cross section containing saturation are employed.
The calculation of the neutrino cross section with the ASW structure
function has also been performed in Ref.~\cite{Machado:2005af}. 
Though the quantitative agreement of our
result with this calculation is reasonably good, some discrepancy appears
due to the fact that to extend the
validity of ASW  to the region $x>0.01$,
we have connected the ASW
structure function to the parameterization of
HERA data {\it \`a} $la$ BCDMS from Ref.~\cite{SMC}.

The effect of a rapid rise of 
the $F_2$ structure function at low $x$ in the CC neutrino-nucleon cross
section has also been studied using the $x$-slope $\lambda(Q^2)$
of the $F_2$ HERA data ($F_2 \sim x^{-\lambda}$) for all $Q^2$ values. 
We have found that the cross section rises with energy very rapidly.
At $E=10^{10}$ GeV it can become a factor 4 above the 
pQCD calculation with parton densities. 
On the other hand the logarithmic rise of the structure function
$F_2$ at small $x$ predicted by the DLA-pQCD results in a slower increase of
the CC neutrino-nucleon cross section with energy. At $E=10^{10}$ GeV
the DLA estimation is a 20 $\%$ below the pQCD calculation with
parton densities. When considering only realistic extrapolations,
the theoretical uncertainty at $E=10^{9}$ GeV is a factor 2.
 
The obtained uncertainty for the tau energy loss is to be implemented,
together with the corresponding one for the CC 
neutrino-nucleon cross section, 
both in analytical and Monte Carlo calculations of the rates
of taus emerging from Earth-skimming tau neutrinos, which is currently being 
used to calculate high energy neutrino bounds. This  
task is beyond the scope of this paper.


%
\vspace{1cm}
\noindent
\Large{} {\bf Acknowledgements}    \vspace{0.5cm}

\normalsize{}
\medio

We thank O. Blanch Bigas, M.V.T. Machado, D. Pertermann, Yu.M. Shabelski,
and D.A. Timashkov for useful comments on this work.
N.A. acknowledges financial support by Ministerio de Educaci\'on y Ciencia
(MEC) of Spain under a Ram\'on y Cajal contract.
This work has been supported by MEC under grants 
FPA2005-01963 and FPA2004-01198,  by Xunta de Galicia under grant
2005 PXIC20604PN and Conseller\'\i a de Educaci\'on, and
by FEDER Funds.

\newpage

\vspace{1cm}
\noindent
\Large{} {\bf Appendix: The ASW $F_2$ 
structure function}    
\vspace{0.5cm}

\normalsize{}
\medio

The form of the  single universal curve
where all data on $\sigma^{\gamma^* p}$ and on
$\sigma^{\gamma^* A}$ lie
as function of the scaling variable $\tau=Q^2/Q^2_{sat}$ is motivated
by saturation and given by~\cite{ASW,Albacete:2005ef}:
\begin{equation}
  \sigma^{\gamma^* p}(x,Q^2) \equiv
  \Phi(\tau) =
\bar\sigma_0
  \left[ \gamma_E + \Gamma\left(0,\xi\right) +
         \ln\xi \right]\, ,
       \label{eqscalf}
\end{equation}
with  $\gamma_E$ the Euler constant, $\Gamma\left(0,\xi\right)$
the incomplete $\Gamma$ function, and $\xi=a/\tau^b$,
with $a=1.868$ and $b=0.746$ extracted from a fit to lepton-proton data.
The saturation scale  $\qs^2$ is parameterized as 
$\qs^2$(GeV$^{2}$) $=(\bar{x}/x_0)^{-\lambda}$~\cite{Golec-Biernat:1998js},
where $x_0= 3.04\cdot 10^{-4}$, $\lambda=0.288$, and $\bar
x=x\,(Q^2+4m_f^2)/Q^2$ with $m_f=0.14$~GeV. 
The normalization is fixed by $\bar\sigma_0=40.56$ $\mu$b.

\begin{figure*}[t]
\includegraphics[width=13.0cm]{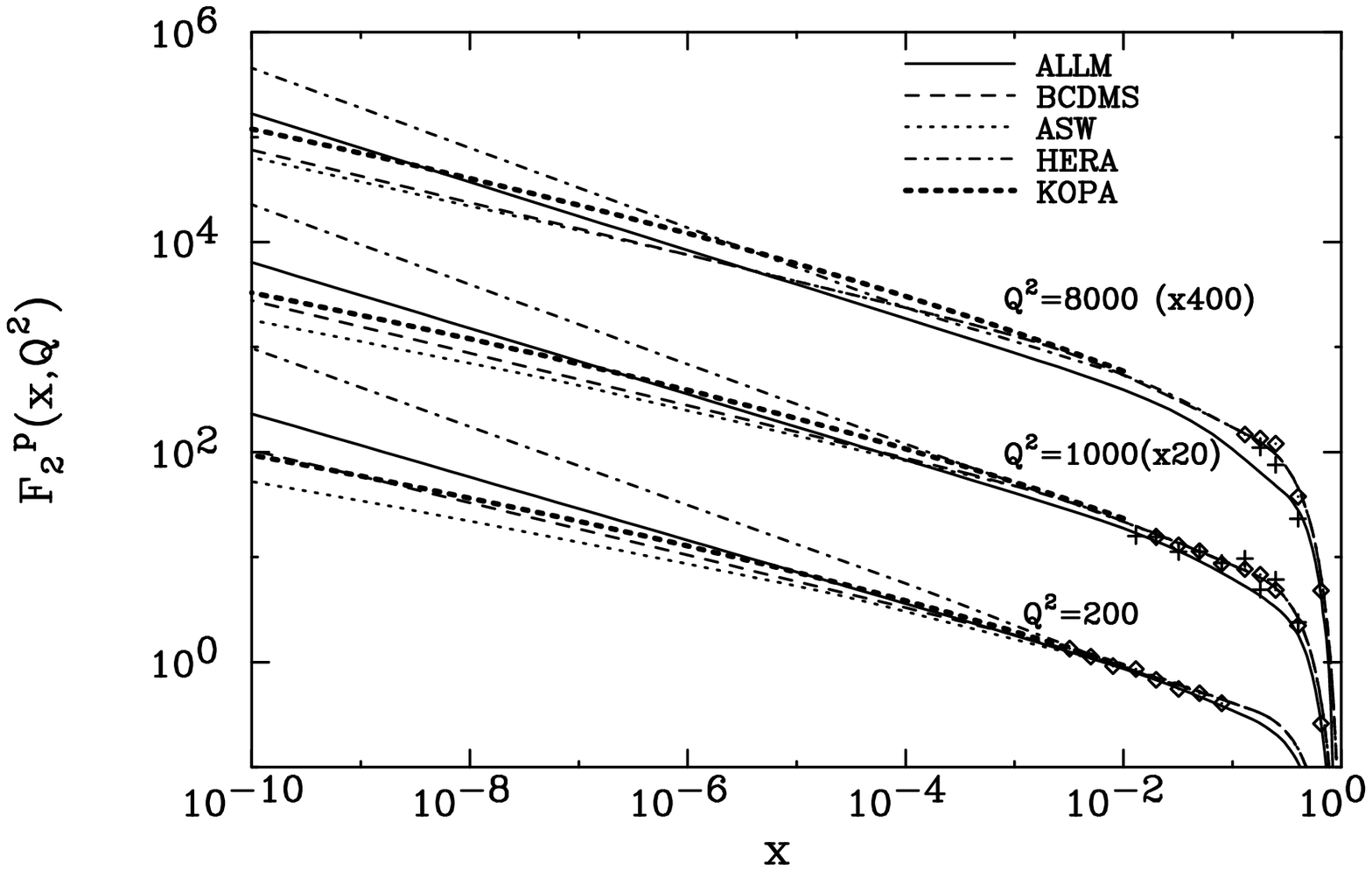}
\caption{The proton structure function $F_2$ as a function
of $x$ for different
high $Q^2$(GeV$^2$) values. Data points are from HERA
\cite{H1DATA,ZEUSDATA}. } 
\label{figF2a}
 \end{figure*}

\begin{figure*}[t]
 \includegraphics[width=13.0cm]{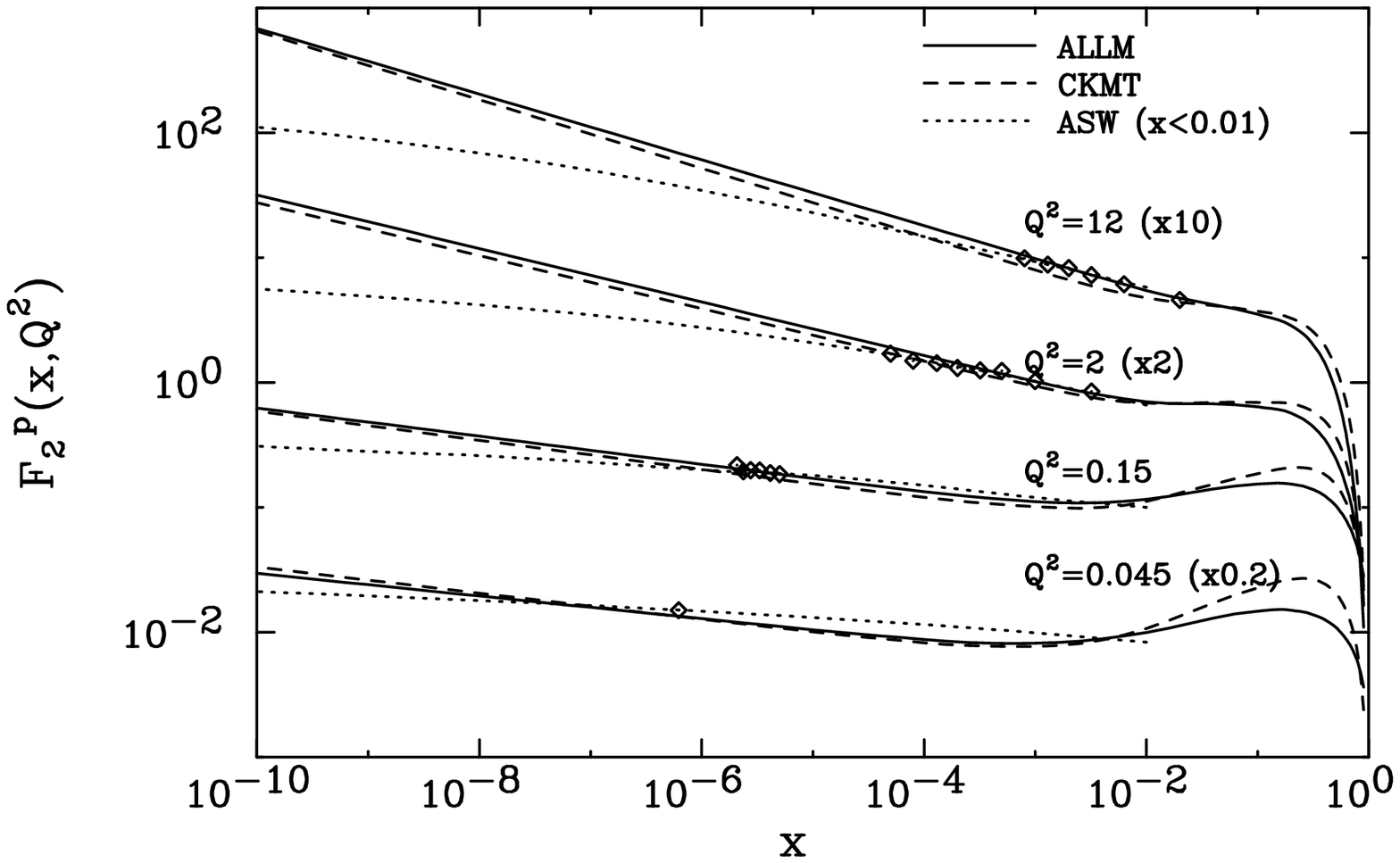}
 \caption{The proton structure function $F_2$ as a function of $x$
for different low $Q^2$ (GeV$^2$) values. Data points are from HERA
\cite{H1DATA,ZEUSDATA}. } 
\label{figF2b}
 \end{figure*}

The extension to the nuclear case is done through
\begin{equation}
\sigma^{\gamma^*A}=\frac{\pi R_A^2}{\pi R_p^2}
\sigma^{\gamma^*p}(\tau_A)
\label{eqnormal}
\end{equation}
and
\begin{equation}
   Q_{\rm sat,A}^2=Q_{\rm sat,p}^2\left(\frac{A \pi R_p^2}
   { \pi R_A^2}\right)^\frac{1}{\delta} \hspace{-0.1cm}
   \Rightarrow
   \tau_A=\tau \; \left[\frac{\pi R_A^2}{A \pi R_p^2}\right]^\frac{1}{\delta},
   \label{eqtaua}
\end{equation}
where the nuclear radius is given by the usual parameterization
$R_A=(1.12 A^{1/3}-0.86 A^{-1/3})$ fm, and $\delta=0.79\pm0.02$ and $\pi
R_p^2=1.55 \pm 0.02$
fm$^2$ are extracted from a fit to lepton-nucleus data.
The nuclear structure function $F_2^A$ is
$F_2^A(x,Q^2)=Q^2 \sigma^{\gamma^*A}/(4\pi^2\alpha)$.
The ASW structure function for the proton case is recovered by taking  $A=1$
in the expressions above (see Figs.~\ref{figF2a} and~\ref{figF2b}).

The functional shape of (\ref{eqscalf})
is motivated by considerations in saturation physics
\cite{ASW,Albacete:2005ef}. From a pragmatic point of view, it provides a very
good description of existing lepton-proton and lepton-nucleus data in the
region $0.01<\tau,\tau_A<100$ and $x<0.01$ which for $Q^2=0.01, 0.1, 1$,
and $10$ GeV$^2$ corresponds to a low $x$ limit of $\sim 10^{-5}$, $10^{-7}$,
$10^{-10}$, and $10^{-13}$, respectively. 
For $\tau\to 0$  $F_2$ behaves like a single
logarithm, so $F_2\propto \ln{1/x}$ for $x\to 0$ and $F_2/A\propto
\ln{A}/A^{1/3}$ for $A\to \infty$. Thus this model results in very large
screening corrections for asymptotic values of $x$ and $A$.



\newpage

\newpage

\newpage

\newpage

\newpage

\vskip-1cm

\newpage

\begin{thebibliography}{99}

\medio
\bibitem{Halzen}
  T.~K.~Gaisser, F.~Halzen and T.~Stanev,
  Phys. Rept.  258 (1995) 173
  [Erratum-ibid. 271 (1996) 355].
\bibitem{Fargion} D.~Fargion,
  Astrophys.\ J.\  570 (2002) 909.
\bibitem{Bertou} X.~Bertou, P.~Billoir, O.~Deligny, C.~Lachaud,
and A.~Letessier-Selvon,
Astropart.\ Phys.\ 17 (2002) 183.
\bibitem{EZas}
  E.~Zas,
  New J. Phys. 7 (2005) 130.
\bibitem{ICRCAuger_neutrino1} O. Blanch Bigas (Pierre Auger Collaboration),
Proceedings of the 30th ICRC (2007), to be published.
\bibitem{ICRCAuger_neutrino2} J. Alvarez-Mu\~niz (Pierre Auger Collaboration),
Proceedings of the 30th ICRC (2007), to be published.
\bibitem{Dutta2001} S.I. Dutta, M.H. Reno, I. Sarcevic, and D. Seckel,
Phys. Rev. D63 (2001) 094020.
\bibitem{Bugaev2004} E.V. Bugaev, T. Montaruli, Yu.V. Shlepin, and
I. Solkalski, Astropart. Phys. 21 (2004)~491.
\bibitem{Aramo2005} C. Aramo {\it et al.}, Astropart. Phys. 23 (2005)~65.
\bibitem{nosoICRC07} N.~Armesto, C.~Merino, G.~Parente, and E.~Zas,
Proceedings of the 30th ICRC (2007), to be published.
\bibitem{nuclear} M. Arneodo, Phys. Rep. 240 (1994) 301;
  D.~F.~Geesaman, K.~Saito, and A.~W.~Thomas,
  Ann.\ Rev.\ Nucl.\ Part.\ Sci.\  45 (1995) 337.
\bibitem{saturacion}
QCD Perspectives on Hot and Dense Matter, edited by J.-P. Blaizot and
E. Iancu, Kluwer, Dordrecht, The Netherlands, 2002 (NATO Science Series, II, Mathematics, Physics, and Chemistry, Vol. 87).

\bibitem{ASW} N. Armesto, C. Salgado, and U.A. Wiedemann, Phys. Rev. Lett. 94
  (2005) 022002.
\bibitem{BM2002} A.V. Butkevich and S.P.Mikheyev, Zh. Eksp. Teor. Fiz 122
  (2002) 17.
\bibitem{BB} L.B. Bezrukov and E.V. Bugaev, Yad. Fiz. 33 (1981) 1195.
\bibitem{BS2003} E.V. Bugaev and Yu.V. Shlepin, Phys. Rev. D67 (2003) 934027.
\bibitem{Lipari} P. Lipari and T. Stanev, Phys. Rev. D 44 (1991) 3543.
\bibitem{KLS2005} K.S. Kuzmin, K.S. Lokhtin, and S.I. Sinegovsky,
  Int. J. Mod. Phys. A20 (2005) 6956;
  A.A.~Kochanov, K.S.~Lokhtin, and S.I.~Sinegovsky,
  arXiv:hep-ph/0508306;
K.S.~Lokhtin and S.I.~Sinegovsky,
  Russ.\ Phys.\ J.\  49 (2006) 326
  [Izv.\ Vuz.\ Fiz.\  49 (2006) 82].
\bibitem{Petrukhin} D.A. Timashkov and A.A. Petrukhin, 
Proceedings of the 29th ICRC (2005) 9, 89-92.
\bibitem{Dutta2005} S.I. Dutta, Y. Huang, and M.H. Reno, Phys. Rev.
D72 (2005) 013005.
\bibitem{ALLM} H. Abramowicz and A. Levy, arXiv:hep-ph/9712415.
\bibitem{CKMT}  A. Capella, A. Kaidalov, C. Merino, and J. Tran
  Thanh Van, Phys. Lett. B337 (1994)~358;
A. Kaidalov, C. Merino, and D. Pertermann, Eur. Phys. J. C20 (2001) 301.
\bibitem{F2petrukhin} A.A. Petrukhin and  D.A. Timashkov,
Yad. Fiz. 67 (2004) 2241 [Phys. At. Nucl. 67 (2004) 2216].
\bibitem{H1DATA} C. Adloff {\it et al.} (H1 Collaboration),
Eur. Phys. J. C21 (2001) 33.
\bibitem{ZEUSDATA} J.~Breitweg {\it et al.} (ZEUS Collaboration), Phys. Lett. B487 (2000) 53.
\bibitem{Stasto} A.M. Stasto, K. Golec-Biernat, and J. Kwiecinski,
  Phys. Rev. Lett. 86 (2001) 596.
%
\bibitem{ReportNestor} N. Armesto, J. Phys. G32 (2006) R367. 
\bibitem{CastroPena}
  J.A.~Castro Pena, G.~Parente, and E.~Zas,
  Phys.\ Lett.\  B507 (2001) 231.
\bibitem{Anchordoqui} 
  L.A.~Anchordoqui, A.M.~Cooper-Sarkar, D.~Hooper, and S.~Sarkar,
  Phys.\ Rev.\  D74 (2006) 043008.

\bibitem{SMC} 
  B.~Adeva {\it et al.}  (Spin Muon Collaboration),
  Phys.\ Rev.\  D58 (1998) 112001.
\bibitem{HERA} 
  C.~Adloff {\it et al.}  (H1 Collaboration),
  Phys.\ Lett.\  B520 (2001) 183.


\bibitem{KOPA} A.V.~Kotikov and G.~Parente,
  Nucl.\ Phys.\  B549 (1999) 242.
%
\bibitem{Kutak_Kwiecinski} K. Kutak and J. Kwiecinski, Eur. Phys. J. C29 (2003) 521.
%
\bibitem{Henley:2005ms}
  E.~M.~Henley and J.~Jalilian-Marian,
  Phys.\ Rev.\  D73 (2006) 094004.
\bibitem{Machado:2005af}
  M.~V.~T.~Machado,
  Phys.\ Rev.\  D71 (2005) 114009.
%
\bibitem{Albacete:2005ef} J.L.~Albacete, N.~Armesto, J.G.~Milhano,
  C.A.~Salgado, and U.A.~Wiedemann,
  Eur.\ Phys.\ J.\  C 43 (2005) 353.
\bibitem{Golec-Biernat:1998js} K.~Golec-Biernat and M.~W\"usthoff,
Phys.\ Rev.\ D 59 (1999) 014017.
%
\end{thebibliography}
\end{document}